\documentclass[12pt]{article}
\usepackage{epsf}
\usepackage{\jobname}
\begin{document}
\title{The heat kernel coefficients for the dielectric cylinder}
\author{
  {\sc M. Bordag}\thanks{e-mail: Michael.Bordag@itp.uni-leipzig.de} \\
   \small  University of Leipzig, Institute for Theoretical Physics\\
  \small  Augustusplatz 10/11, 04109 Leipzig, Germany\\
  and
  {\sc I. G. Pirozhenko}\thanks{e-mail: pirozhen@thsun1.jinr.ru}\\
  \small Bogolyubov Laboratory of Theoretical Physics,\\
  \small Joint Institute for Nuclear Research, 141980 Dubna, Russia}
\maketitle
\begin{abstract}
 We calculate the \hkks for the \elm field in the background of a dielectric
 cylinder with non equal speeds of light inside and outside. The coefficient
 $a_{2}$ whose vanishing makes the vacuum energy of a massless field unique,
 turns out to be zero in dilute order, i.e., in order $(\ep-1)^{2}$, and
 nonzero beyond. As a consequence, the vanishing of the vacuum energy in the
 presence of a dielectric cylinder found by Casimir-Polder summation must take
 place irrespectively of the methods by which it might be calculated.
\end{abstract}
\thispagestyle{empty}
\section{Introduction}\label{Sec1}
The vacuum energy of the \elm field in the presence of a dielectric body is a
topic of considerable interest since Schwingers attempt to explain
sonoluminescence. Although this seems unlikely to be possible, mainly due to
different time scales involved, open questions connected with the
renormalization remain.

The divergent part of the vacuum energy can be expressed in terms of the
corresponding \hkksp By means of the renormalization procedure the divergent
part must be subtracted from the \gse and added as counter term to the
classical energy of the background. This general procedure does not work
straightforwardly for a dielectric body. The point is that there is no
classical energy which may be associated with it. Note that there are no
electric or magnetic fields present, instead there is merely the ability of
the body to get polarized upon application of an \elm field. One may overcome
this problem by introducing formally the necessary terms for a classical
energy and put zero the coefficients in front of them. This is similar to
semiclassical gravity where one assumes the coefficients in front of the terms
quadratic in curvature to be sufficiently small. But there is one more problem
associated with the vacuum energy of a dielectric body. As the field is
massless, there is no normalization condition to get a unique \gse as long as
the heat kernel coefficient $a_{2}$ is nonzero because it comes with the
logarithmic divergence, see \cite{Bordag:1999vs} for a detailed explanation.

In \cite{Bordag:1999vs} the \hkks for a generic smooth dielectric background
and for a dielectric ball have been calculated. The coefficient $a_{2}$ turned
out to be nonzero in general. For small deviation of the dielectricity $\ep$
from unity (resp. from its value at spatial infinity) $a_{2}$ is zero to first
order in $(\ep-1)$ and nonzero starting from second order, $(\ep-1)^{2}$,
which is called dilute approximation. For the dielectric ball it is zero in
second order in addition. In this way, for a more singular background
configuration ($\ep(r)=\ep\Theta(R-r)$ has a jump) the singularity is slightly
weaker. It is this zero of $a_{2}$ to order $(\ep-1)^{2}$ which allowed to get
a unique result for the vacuum energy of the dielectric ball in dilute
approximation,
\be\label{dilball0}\E_{0}^{\rm ball}={23\over 1536\pi}{(\ep-1)^{2} \over R}
+O\left((\ep-1)^{3}\right),
\ee
which was obtained in by Casimir-Polder summation in \cite{Milton:1997ky}, by
a perturbative setup in \cite{Barton:1999ic} and by mode summation in
\cite{Nesterenko:2000cf}.

In the present paper we consider a dielectric cylinder and calculate the \hkks
$a_{n}$ for $n\le2$. This problem is technically more involved than the
corresponding calculation for the ball. Although the Jost function of the
corresponding scattering problem which we use is expressed in terms of Bessel
functions the photon polarizations do not separate and there is an additional
dependence on the axial momentum.

The vacuum energy of a dielectric cylinder had been calculated in
\cite{Milton2000} in dilute approximation by summing the Casimir-Polder
potentials and it turned out to be zero,
\be\label{dilball}\E_{0}^{\rm cyl}=O\left((\ep-1)^{3}\right).
\ee
In the present paper we show that the corresponding $a_{2}$ is zero in this
(dilute) approximation and conclude that the vacuum energy in dilute order is
zero irrespectively of the method by which it is calculated.

In this paper we use units with $\hbar=c=1$.

\section{Basic formulas}
In order to regularize the vacuum energy of the \elm field we use the
zeta functional regularization and express the energy in terms of the
corresponding zeta function,
\be\label{E0}
\E_{0}(s)=\frac{\mu^{2}}{2} \ \zeta\left(s-\frac12\right).
\ee
In the presence of the dielectric cylinder the spectrum of the \elm
field is completely continuous and we use the representation of the
regularized \gse derived in \cite{Bordag:1996fv} for a spherically symmetric
background which was in \cite{Bordag:1998tg} rewritten for a cylindrically
symmetric background. After dropping the Minkowski space contribution
we obtain
\be\label{zeta}\zeta(s)=\frac{\sin(\pi s)}{\pi}
\sum_{l=-\infty}^{\infty}
\int\limits_{0}^{\infty}\frac{d k_z}{\pi}
\int\limits_{k_z}^{\infty} d k\,(k^2-k_z^2)^{-s}
\frac{\partial}{\partial k}\ln \Delta_{l}(k,k_{z}).
\ee
Here, $\Delta_{l}(k,k_{z})$ is the Jost function for the scattering of \elm
waves off the cylinder with radius $R$  on the imaginary axis. It is given by
\bea\label{jostfct}
\Delta_{l}(k,k_{z})&=&
\frac{1}{\Delta^{\infty}_l}
\left\{\Delta_{l}^{TM}(k,k_{z})\Delta_{l}^{TE}(k,k_{z})\right. \\ && \left. 
+l^2\, (k^{2}-k_{z}^{2})\,k_{z}^{2}\,(1-\al^{2})^{2}
 c_{2}^{-2} \left[I_{l}(qR)K_{l}(kR)\right]^{2}\right\} \nn
\eea
with
\bea\label{tem}
\Delta_{l}^{TM}(k,k_{z})&=&
\mu_1\,k\,I'_{l}(qR)\,K_{l}(kR)-\mu_2\,q\,I_{l}(qR)\,K'_{l}(kR),\nn \\
\Delta_{l}^{TE}(k,k_{z})&=&
\ep_1\,k\,I'_{l}(qR)\,K_{l}(kR)-\ep_2\,q\,I_{l}(qR)\,K'_{l}(kR),\nn \\
\Delta^{\infty}_l(k,k_z)&=&\frac{1}{4}\,e^{2\,(q-k)}\, q k \,(\varepsilon_1 k+
\varepsilon_2 q)\,(\mu_1 k+\mu_2 q),
\eea
and the notations
$q=\sqrt{\alpha^2\,k^2+k_z^2\,(1-\alpha^2)}$, $\al\equiv c_{2}/c_{1}$;
and the speeds of light inside and outside are $c_{i}=1/\sqrt{\ep_{i}
  \mu_{i}}$ ($i=1,2$). This formula can be found in textbooks on classical
electrodynamics, see also \cite{Nesterenko:1999zg}.

The heat kernel coefficients are known to be defined by the 
residues of the  zeta function  at the corresponding points $s$,
\be\label{an}a_{n}=(4\pi)^{3/2} \res_{s=\frac32-n} \Gamma(s)\zeta(s).
\ee
These residua result from large all, the momentum $k$, the axial
momentum $k_{z}$ and the orbital momentum $l$. To factorize the orbital
momentum dependence for the terms with $l\ne0$ 
we use the uniform asymptotic expansion of the Bessel functions~\cite{abra70b}.
Being  inserted into Eq. \Ref{jostfct} it
gives raise to the expansion
\be\label{lnas}\ln \Delta_{l}(lk,lk_{z})\sim
\sum_{i=-1,0,1,\dots}{D_{i}(k,k_{z})\over l^{i}}
\ee
for $k,l\to\infty$. Substituting  the latter into
\Ref{zeta}  previously changing the variables
$k\to lk$ and $k_{z}\to l k_{z}$ 
so that $k$ and $k_{z}$ were fixed,
we split the zeta function into  parts according to
\begin{equation}
\Gamma(s) \,\zeta_{cyl}(s)=\tilde{A}(s)\,
+\,2\,\left\{ A_{-1}(s)
+ A_0(s)+A_1(s)+ A_2(s)+ A_3(s)+\dots
\right\}.
\label{e9}
\end{equation}
Here $\tilde A(s)$ results from $l=0$
\be
\tilde{A}(s)=\frac{R^{2s-1}}{c_2^{2s}\,\Gamma(1-s)}\!
\int\limits_{0}^{\infty}\frac{dk_z}{\pi}\!
\int\limits_{k_z}^{\infty}\! dk (k^2-k_z^2)^{-s}
\frac{\partial}{\partial k}\ln {\Delta_0}(k,k_z),
\label{zeroterm}
\ee
and below we will use the asymptotic expansion of the Bessel functions
for large argument in it. The $A_i(s)$'s are generated by the
expansion \Ref{lnas}
\begin{equation}
A_i(s)=\frac{R^{2s-1}\zeta_{R}(2s-1+i)}{c_2^{2s}\,\Gamma(1-s)}\!
\int\limits_{0}^{\infty}\frac{dk_z}{\pi}\!
\int\limits_{k_z}^{\infty}\! dk (k^2-k_z^2)^{-s}
\frac{\partial}{\partial k}\ln D_i(k,k_{z}).
\label{e7}
\end{equation}
For the heat kernel coefficients up to $a_2$ it is sufficient to include $i$
up to $3$. The summation over $l$ (from 1 to $\infty$) was  carried out in 
$A_i(s)$ and resulted in the Riemann zeta function.

The expressions for the $D_{i}$ are simple for $i=-1$ and $i=0$, but
sufficiently involved for $i=1,2,3$ to be banned into the
appendix. Here we note
\begin{eqnarray}\label{e6}
D_{-1}&=&2 \left(q-k+\sqrt{1+q^2}-\sqrt{1+k^2}+
\ln\frac{q}{k}\,\frac{1+\sqrt{1+q^2}}{1+\sqrt{1+k^2}}\right),\\
D_0&=&\ln\left\{\left[\mu_1\,\frac{k}{q}\left(\frac{1+q^2}{1+k^2}
\right)^{1/4}+
\mu_2\,\frac{q}{k}\left(\frac{1+k^2}{1+q^2}\right)^{1/4}\right]
\times\Biggl[\mu_i\leftrightarrow\varepsilon_i\Biggr]
\right.\nonumber\\
&&\left.
+\frac{(k^2-k_z^2)\,
(1-\alpha^2)^2\,k_z^2}{c_2^2\,k^2\,q^2\,\sqrt{1+q^2}\,
\sqrt{1+k^2}}
\right\}+\ln\left\{\frac{q \,k}{(\varepsilon_1 k+\varepsilon_2 q)
(\mu k+\mu_2 q)}
\right\}.\nn
\end{eqnarray}

In the next section the functions $A_i\; (i=-1,..3)$ and
$\tilde{A}(s)$ are considered one after another in order to find their
residues contributing to the heat-kernel coefficients.
%
\section{Calculation of the heat-kernel coefficients}

In this section we give a detailed calculation of the heat-kernel coefficient
$a_2$.  The results of the analogous calculations for the junior coefficients
$a_n$ with $n<2$ are listed in the end of the section.

We start from $A_{-1}(s)$, where it is easy to carry out the integration 
explicitly 
\begin{equation}
A_{-1}(s)=-\frac{R^{2 s-1}\zeta(2s-2)}{c^{2s}_2\,\pi}(\alpha^{2 s}-1)
\frac{\Gamma(s-1)}{1-2s}.
\label{e10}
\end{equation}
We are interested in the residue of $A_{-1}(s)$ at the point
$s=-1/2$. As $A_{-1}$ has no pole at this point, the residue equals
zero. So, this term of the expansion~(\ref{e9}) doesn't contribute to
$a_2$.

The term $A_0(s)$ is more complicated. To simplify the integration  we make
the change of variables $k\to k,\;k_z\to k \eta$. It gives
\beao
A_{0}(s)&=&\frac{\zeta(2s-1)}{\pi}\frac{R^{2s-1}}{c_2^{2s}\,\Gamma(1-s)}\, \\
&&\int\limits_{0}^{1}d\eta\,(1-\eta^2)^{-s}
\!\int\limits_{0}^{\infty}\! dk\, k^{-2s+1}
\left[\frac{\partial}{\partial k}-\frac{\eta}{k}
\frac{\partial}{\partial \eta}\right] D_{0}(k,\eta).
\eeao
The function 
\begin{eqnarray}
D_{0}(k,\eta)&=&
\ln\left\{\left[\mu_1\,\left(\frac{1+
\gamma^2 k^2}{1+k^2}\right)^{1/4}+\mu_2\,
\gamma^2\left(\frac{1+k^2}{1+\gamma^2 k^2}\right)^{1/4}
\right]
\times\Biggl[\mu_i\leftrightarrow\varepsilon_i\Biggr]
\right.\nonumber\\
&&+\left.\frac{(1-\eta^2)\,
(1-\alpha^2)^2\,\eta^2}{c_2^2\sqrt{1+\gamma^2\,k^2}\,
\sqrt{1+k^2}}
\right\}+\ln \frac{1}{\gamma\,(\varepsilon_1+
\varepsilon_2\,\gamma)(\mu_1+\mu_2\,\gamma)},\nonumber
\end{eqnarray}
with $\gamma=\sqrt{\alpha^2+(1-\alpha^2)\,\eta^2}$,
has the following asymptotics at zero and infinity,
\beao
D_0(k,\eta)|_{k\to0}&=&\ln\frac{(\mu_1+\gamma^2\mu_2)(\varepsilon_1+\gamma^2 \varepsilon_2)+(1-\eta^2)\, 
\eta^2\,(1-\alpha^2)^2\, c_2^{-2}}{\gamma 
(\varepsilon_1+\varepsilon_2 \gamma)(\mu_1+\mu_2 \gamma)}\\&&+{\cal{O}}(k^2), 
\nn\\
D_0(k,\eta)|_{k\to\infty}&=&\frac{(1-\alpha^2)^2 \eta^2 (1-\eta^2)}{2 c_2^2 
\gamma^2 (\mu_1+\mu_2 \gamma)
(\varepsilon_1+\varepsilon_2 \gamma)}\,\frac{1}{k^2}+
{\cal{O}}\left(k^{-3}\right).
\eeao
Therefore the integral over $k$ in $A_{0}$ converges in 
the range $-1/2<s<1/2$.
 Adding and subtracting from the 
integrand its asymptotics at $k\to\infty$ we obtain the analytic 
continuation into the vicinity of the point $s=-1/2$
\bea
A_{0}(s)&=&\frac{\zeta(2s-1)}{\pi}
\frac{R^{2s-1}}{c_2^{2s}\,\Gamma(1-s)} 
\int\limits_{0}^{1}d\eta\,(1-\eta^2)^{-s} 
\label{e12}                    \\  &&  \hspace{-1.5cm}
\times\Biggl\{\int\limits_{0}^{\infty}\!
\frac{ dk}{k^{2s-1}}
\left[\frac{\partial}{\partial k}-\frac{\eta}{k}
\frac{\partial}{\partial \eta}\right]
\left[D_{0}(k,\eta)-
\frac{\eta^2
(1-\eta^2)(1-\alpha^2)^2}{2\,c_2^2 \gamma^2 (\mu_1+\mu_2 \gamma)
(\varepsilon_1+\varepsilon_2 \gamma)}\frac{1}{(k+1)^2}
\right]\Biggr.  \nn \\ && \hspace{-1.5cm}
-\Biggl.\frac{1}{2}\left[(-2s+1)+\eta
\frac{\partial}{\partial \eta}\right] \Gamma(-2s+1)
\Gamma(2s+1)
\frac{\eta^2
(1-\eta^2)(1-\alpha^2)^2}{c_2^2 \gamma^2 (\mu_1+\mu_2 \gamma)
(\varepsilon_1+\varepsilon_2 \gamma)}\Biggr\}.  \nn
\eea

The first integral is now analytical around $s=-1/2$, 
the second one contains the
pole. But this pole is canceled by the Riemann zeta function $\zeta(2s-1)$
which equals zero at $s=-1/2$. As a result we have
\begin{equation}
\mathop{res}\limits_{s\to-1/2} A_{0}^{sing}(s)=0.
\label{e13}
\end{equation}

Analyzing now the asymptotics of $D_{1}(s)$ at zero and at infinity,
$$
\bigl.D_1\bigr|_{k\to0}\to\frac{1}{2}(\gamma^2-1)\,k^2+{\mathcal O}(k^4),
\;\;\;
\bigl.D_1\bigr|_{k\to\infty}\to\frac{\Theta_1}{k}+{\mathcal O}(k^{-3}),
$$ 
where
$$
\Theta_1=\frac{1}{4}\left\{3-\frac{3}{\gamma}+2\,\frac{\varepsilon_2-
\varepsilon_1}{\varepsilon_1+\varepsilon_2 \gamma}+\frac{\mu_2-
\mu_1}{\mu_1+\mu_2 \gamma}\right\}, 
$$
one comes to the conclusion that the integral over $k$
in
\begin{equation}
A_{1}(s)=\frac{\zeta(2s)}{\pi}\frac{R^{2s-1}}{c_2^{2s}\,\Gamma(1-s)}
\!
\int\limits_{0}^{1}d\eta\,(1-\eta^2)^{-s}
\!\int\limits_{0}^{\infty}\! dk\, k^{-2s+1}
\left[\frac{\partial}{\partial k}-\frac{\eta}{k}
\frac{\partial}{\partial \eta}\right]D_{1}(k,\eta),
\label{e18}
\end{equation}
converges in the strip $0<s<3/2$.
The analytic continuation to the point $s=-1/2$ is the following
\bea\label{e19} 
 A_{1}(s)&=&\zeta(2s)\,\frac{c_2^{-2s}\,R^{2s-1}}{\pi \, \Gamma(1-s)}
 \\
&& \times \left\{   \int\limits_{0}^{1}\frac{d\eta}{(1-\eta^2)^{s}}
\!\int\limits_{0}^{\infty}\! dk\, k^{-2s+1}
\left\{\frac{\partial}{\partial k}-\frac{\eta}{k}
\frac{\partial}{\partial \eta}\right\}\left[D_{1}(k,\eta)-
\frac{\Theta_1(\eta)}{\sqrt{k^2+1}}\right]  \right.          \nonumber\\  
  &&\left.
-\Gamma(s)
\int\limits_{0}^{1}\frac{d\eta}{(1-\eta^2)^{s}}\left[\Gamma(-s+3/2)+
\frac{\Gamma(-s+1/2)}{2}\,\eta \, \frac{\partial}{\partial \eta
}\right]\, \Theta_1(\eta)   \right\}  . \nn
\eea
Here the asymptotics of $D_1(k,\eta)$ for $k\to\infty$ was added and 
subtracted.
Both terms in (\ref{e19}) are analytic around $s=-1/2$. Therefore
$\mathop{res}\limits_{s\to-1/2} A_{1}^{sing}(s)
=0$.

Similar arguments for $A_{2}(s)$, where the integral with respect to
$k$ exists at $-1/2<s<3/2$,
give
\bea\label{e21}
\mathop{res}\limits_{s\to-1/2}^{} A_2(s)&=&
-\frac{R^{-2}}{32}\frac{c_2}{\pi^{3/2}} 
\int\limits_0^1 \frac{d\eta}{\sqrt{1-\eta^2}} \\ &&
\times \left[\frac{3\mu_2^2\gamma^4-\mu_2^2\gamma^2-4\gamma^2\mu_1\mu_2+
3\mu_1^2-\gamma^2\mu_1^2}{\gamma^2\,(\mu_1+\mu_2)^2}+
(\mu\leftrightarrow\varepsilon)\right]. \nn
\eea

The next term in (\ref{e9}) we have to consider is $A_3(s)$.
The integrals in $A_3(s)$ converge at $s=-1/2$, while the Riemann zeta 
function has a pole at this point. Thus
\begin{equation}
\mathop{res}\limits_{s\to-1/2}^{} A_3(s)=
\frac{R^{-2}\,c_2}{\pi^{3/2}}
\int\limits_0^1 d\eta \,\sqrt{1-\eta^2} \int\limits_0^{\infty}dk\,k^2
\left[\frac{\partial}{\partial k}-\frac{\eta}{k}\frac{\partial}{\partial\eta}\right]D_3.
\label{e22}
\end{equation}

Now we are left with the term $\tilde{A}(s)$ in Eq. (\ref{zeroterm}). 
\begin{equation}
\tilde{A}(s)=\frac{R^{2s-1}}{c_2^{2s}\,\pi\,\Gamma(1-s)}\,
\int\limits_{0}^{1}d\eta\,(1-\eta^2)^{-s}
\!\int\limits_{0}^{\infty}\! dk\, k^{-2s+1}
\left[\frac{\partial}{\partial k}-\frac{\eta}{k}
\frac{\partial}{\partial \eta}\right] \ln{\Delta_0}(k,\eta).
\label{tilde_a}
\end{equation}
The asymptotic behavior
of $\ln{\Delta_0}(k,\eta)$ at $k=0$ and $k=\infty$ is the following
\begin{eqnarray}
\ln\tilde{\Delta}(k,\eta)|_{k=0}&=&\ln\frac{4 \varepsilon_2 \mu_2
\gamma^3}{(\varepsilon_1+\varepsilon_2 \,\gamma)\,(\mu_1+\mu_2 \,\gamma)}
+2\,(1-\gamma)\,k+{\cal{O}}(k^2),\nonumber\\
\ln\tilde{\Delta}(k,\eta)|_{k=\infty}&=&\frac{1}{8}\,
\frac{-\varepsilon_1 \,\gamma-3 \,\varepsilon_1+3\, \varepsilon_2 \,
\gamma^2 +
\varepsilon_2 \,\gamma}{(\varepsilon_1+\varepsilon_2 \,\gamma)\,\gamma}\,
\frac{1}{k}\label{ass_zer}\\
&&+\frac{-3 \,\varepsilon_2^2 \,\gamma^4+
4 \,\varepsilon_1 \,\varepsilon_2 \,\gamma^2+\varepsilon_2^2 \,\gamma^2+
\varepsilon_1^2 \, \gamma^2-3 \varepsilon_1^2}{16\,\gamma^2 \,
(\varepsilon_1+
\varepsilon_2 \,\gamma )^2 }\,\frac{1}{k^2} \nn \\ && +\{\varepsilon_i
\leftrightarrow\mu_i\}+{\cal{O}}(k^{-3}).\nonumber
\end{eqnarray}
The integrals over $k$ for the first an the second terms in the square
brackets of (\ref{tilde_a}) exist correspondingly in the ranges
$0<s<1$ and $0<s<1/2$.  We divide the integration with respect to $k$
in two parts $\int_0^{\infty}=\int_0^{1}+ \int_1^{\infty}$, where the
first one is regular at $s=-1/2$, and the second one diverges in the
upper bound.  Keeping in mind that only the singular part gives
contribution to the residue we omit the term convergent at $s=-1/2$ .
To obtain the analytic continuation to the point $s=-1/2$ one should
add and subtract from $\ln\tilde{\Delta}(k,\eta)$ two terms of its
expansion in infinity.  After that it is easy to find the residue
\bea\label{tila}
\mathop{res}\limits_{s\to-1/2}^{} \tilde A(s)
&=&\frac{c_2\,R^{-2}}{16\,\pi^{3/2}}
\int\limits_0^1 \frac{d\eta}{\sqrt{1-\eta^2}} \\ && \times
\left[\frac{3\mu_2^2\gamma^4-\mu_2^2\gamma^2-4\gamma^2\mu_1\mu_2+
3\mu_1^2-\gamma^2\mu_1^2}{\gamma^2\,(\mu_1+\mu_2 \gamma)^2}+
(\mu\leftrightarrow\varepsilon)\right], \nn
\eea
which up to the multiplier $-1/2$ coincides with (\ref{e21})

Summing up we see that only the functions $A_2,\;A_3$ and $\tilde{A}$
have nonzero residues in $s=-\frac12$. But as the residues of $A_2$
and $\tilde{A}$ are mutually canceled in (\ref{e9}), the heat-kernel
coefficient $a_2$ is totally defined by the contribution from $A_3$.
The integration over $\eta$ and $k$ in terms of elementary functions
seems impossible in (\ref{e22}). For $\mu_i=1$ simplifications occur,
but still we are able to analyze the behavior of the coefficient $a_2$
as a function of the velocities if light only numerically (see Fig.1).

In the limit of small differences of the velocities of light
(dilute dielectric cylinder) one might expand $\mathop{res}
\limits_{s\to-1/2}^{} A_3(s)$ in powers 
of $c_{1}-c_{2}$. The expansion starts
from the third order
\be
\mathop{res}\limits_{s\to-1/2}^{}\zeta_{cyl}(s)=-\frac{9}{1408}
\frac{R^{-2}}{\sqrt{\pi} c_2^2}(c_1-c_2)^3+\dots .
\ee
The heat kernel coefficient $a_2$ is
\be
a_2=-\frac{\pi}{R^2 c_2^2}\frac{9}{88}(c_1-c_2)^3+\dots .
\label{hk_a2}
\ee
\begin{figure}[t]\unitlength=1cm
\begin{picture}(5,5.5)
\put(-2,-18){\epsfxsize=18cm 
\epsffile{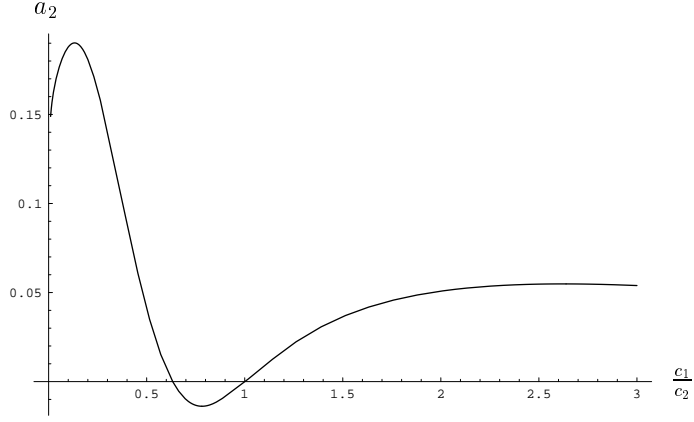}}
\end{picture}
\caption{The \hkk $a_{2}$ for the dielectric cylinder as function of the ration $c_{1}/c_{2}$ of the speeds of light inside vs. outside the cylinder}
\label{figure1}
\end{figure}

The heat-kernel coefficients with $n<2$ are obtained in the same way. For
arbitrary velocities of light inside and outside the cylinder one gets
\beao
a_0&=&-\frac{4\pi\,R^2}{c^3_2}\,(1-\alpha^3),\\
a_{1/2}&=&-\frac{2\,\pi^{3/2}R}{c_2^4}\,\frac{(1-\alpha^2)^2}{
(\mu_1+\mu_2)(\varepsilon_1+\varepsilon_2)},\\
a_1&=&-\frac{2\,\pi}{3 \,c_2 }\,(1-\alpha)
-\frac{8}{c_2}\,\int\limits_0^{\infty}\frac{dk}{k}
\int\limits_0^1\,d\eta
\frac{\eta}{\sqrt{1-\eta^2}}\frac{\partial}{\partial 
\eta} \,D_1(k,\eta),\\
&&+\frac{4}{c_2}\,\int\limits_0^1
\frac{d \eta\,\eta}{\sqrt{1-\eta^2}}\,\\
&& \times \frac{\partial}{\partial \eta}\,\ln\left[\frac{4\,\gamma^4\,
\varepsilon_2\,\mu_2}{(\mu_1+\mu_2\,\gamma^2)(\varepsilon_1+
\varepsilon_2\,\gamma^2)+(1-\eta^2)\,\eta^2\,(1-\alpha^2)^2 \,c_2^{-2}}
\right],\nn\\
a_{3/2}&=&\frac{3 \,\pi^{3/2}}{8\,R}\,\frac{\varepsilon_1^2 \mu_1^2-
4 \varepsilon_1 \mu_1 \varepsilon_2 \mu_2+\varepsilon_2^2\mu_2^2+
\varepsilon_1^2\mu_2^2+\varepsilon_2^2\mu_1^2}{(\varepsilon_1+
\varepsilon_2)^2\,(\mu_1+\mu_2)^2}.
\eeao
%

\section{Conclusions}\label{Secend}
In the preceding sections we calculated the \hkks $a_{n}$ with $n\le2$ of the
\elm field in the background of a dielectric cylinder for different speeds of
light inside and outside. The coefficient $a_2$ vanishes  in the dilute order.

The result of a non vanishing $a_{2}$ beyond the dilute order, obtained in
\cite{Bordag:1999vs} for a dielectric ball and here by Eq. \Ref{hk_a2} for a
cylinder implies an yet unsolved problem. On the one hand side the vacuum
energy of the \elm field in the background of a dielectric body is a physical
quantity and at least in principle measurable. On the other hand side  even
after removing the \uv divergences by hand, it is not uniquely defined due to
the arbitrariness resulting from the logarithmic divergence. The only
explanation which seems meaningful states that the problem itself is ill
defined. The setup of a material body characterized by a position dependent
$\ep(\vec{x})$, a step-like $\ep(r)=\ep\Theta(R-r)$ for instance, is clearly an
idealization. Another example for an idealization are conductor \bc (plane
mirrors or a infinitely thin spherical shell for example) delivering a
uniquely defined vacuum energy.

So this idealization does not work for a dielectric body. This is
surprising in view of the whole area of \elm phenomena like scattering
of waves not connected with vacuum energy where it works well. In
order to attack this problem one has probably to take some microscopic
model for the dielectric body, a lattice of harmonic oscillators for
instance and to calculate the vacuum energy in that background beyond
the dilute approximation. Then one had to investigate a limiting
procedure to turn to the continuous dielectric body. One should expect
that a dependence of the vacuum energy on some parameter like lattice
spacing or plasma frequency remains resulting in the corresponding
logarithmic term connected with $a_{2}$. There are  recent
attempts to do so, see \cite{Barton:2001,Marachevsky:2000yi,Marachevsky:2001pc}.

\section*{Acknowledgments}
The work was supported by the Heisenberg-Landau Program 
and Russian Foundation for Basic Research (Grant No. 00-01-00300)
\section*{Appendix A}
In this appendix we give a list of functions entering the
expansion~(\ref{lnas}).  They are written in terms of the variables
$k$ and $\eta$ with making use of the notations
$t_1=1/\sqrt{1+\gamma^2\,k^2},\;\;
t_2=1/\sqrt{1+k^2},\;\;\gamma=\sqrt{\alpha^2+(1-\alpha^2)\,\eta^2 }$
and

\bea
D_1(k,\eta)&=&{\cal F}_1, \nn\\
D_2(k,\eta)&=&-\,\frac{1}{2}\,{\cal F}_1^2+{\cal F}_2, \nn\\
D_3(k,\eta)&=&{\cal F}_3-{\cal F }_1\times{\cal F}_2,
\eea
where
\begin{eqnarray}
{\cal F}_1&=&\frac{1}{{\cal F}} \left\{\frac{(1-t_1^2)\,
(1-t_2^2)}{t_1^2\,t_2^2}
\left[\varepsilon_1 \mu_1\,
\frac{t_1\,(1-t_2^2)}{t_2\,(1-t_1^2)}\, \left(\frac{5}{12}\,
t_2^3-\frac{1}{4}\,t_2-\frac{3}{4}\,t_1+\frac{7}{12}\,t_1^3
\right)\right.\right.\nn\\
&&+
(\varepsilon_1 \mu_2+\varepsilon_2 \mu_1)\, \left(
-\frac{t_2^3}{12}+\frac{t_2}{4}-\frac{t_1}{4}+
\frac{t_1^3}{12}\right)\nn\\
&&+\varepsilon_2\,\mu_2 \frac{t_2\,(1-t_1^2)}{t_1\,(1-t_2^2)}
\left.\left(\frac{3}{4}\,t_2-\frac{7}{12}\,t_2^3+\frac{t_1}{4}
\frac{5}{12} \,t_1^3 \right)\right]\nn\\ 
&&\left.
+\varepsilon_2 \,\mu_2\,(1-\eta^2)\,\eta^2\,(1-\alpha^2)^2
\,t_1\,t_2\,
\left(\frac{5}{12}\,t_2^3-\frac{t_2}{4}+\frac{t_1}{4}\,-
\frac{5}{12}\,t_1^3\right)\right\},
\end{eqnarray}

\begin{eqnarray}
{\cal F}_2&=&\frac{1}{{\cal F}} \left\{\frac{(1-t_1^2)\,
(1-t_2^2)}{t_1^2\,t_2^2}
\left[\varepsilon_1 \mu_1\,
\frac{t_1\,(1-t_2^2)}{t_2\,(1-t_1^2)}\, \left(\frac{3}{16}\,
t_1\,t_2-\frac{5}{16}t_1\,t_2^3-\frac{41}{48}\,t_2^4
\right.\right.\right.\nn\\
&&\left.-\frac{7}{48}
\,t_1^3\,t_2+\frac{35}{144}\,t_1^3\,t_2^3+\frac{5}{32}\,t_2^2+
\frac{205}{288}\,t_2^6+\frac{13}{16}\,t_1^4-\frac{203}{288}\,
t_1^6-\frac{3}{32}\,t_1^2
\right)\nn\\
&&+
(\varepsilon_1 \mu_2+\varepsilon_2 \mu_1)\, \left(
-\frac{1}{16}\,t_1\,t_2+\frac{1}{48}\,t_1\,t_2^3+\frac{11}{48}\,
t_2^4+\frac{1}{48}\,t_1^3\,t_2-\frac{1}{144}\,t_1^3 \,t_2^3 
\right.\nn\\
&&-
\left.\frac{3}{32}\,t_2^2-\frac{35}{288}\,t_2^6+
\frac{11}{48}\,t_1^4
-\frac{35}{288}\,t_1^6-
\frac{3}{32}\,t_1^2 \right)\nn\\
&&
+\varepsilon_2\,\mu_2 \frac{t_2\,(1-t_1^2)}{t_1\,(1-t_2^2)}
\left(\frac{3}{16}\,t_1\,t_2-\frac{7}{48}\,t1\,t_2^3+
\frac{13}{16}\,t_2^4-
\frac{5}{16} \,t_1^3\,t_2+\frac{35}{144}\,t_1^3\,t_2^3\right.
\nn\\
&&\left.\left.-\frac{3}{32}\,t_2^2-\frac{203}{288}\,t_2^6-
\frac{41}{48}\,t_1^4+\frac{205}{288}\,t_1^6 +
\frac{5}{32}\,t_1^2\right)\right] \nn \\
&&+\varepsilon_2 \,\mu_2\,(1-\eta^2)\,\eta^2\,(1-\alpha^2)^2
\,t_1\,t_2\,
\left(-\frac{1}{16}\,t_1\,t_2+\frac{5}{48}\,t_1\,t_2^2
-\frac{41}{48}\,t_2^4\right. \nn \\ 
&&+\frac{5}{48}\,t_1^3\,t_2
-\frac{25}{144}\,t_1^3\,t_2^3+\frac{5}{32}\,t_2^2+
\frac{205}{288}\,t_2^6-\frac{41}{48}\,t_1^4+
\frac{205}{288}\,t_1^6 \nn\\ 
&&\left.\left.+\frac{5}{32}\,t_1^2\right)\right\},
\end{eqnarray}

\begin{eqnarray}
{\cal F}_3&=&\frac{1}{{\cal F}} \left\{\frac{(1-t_1^2)\,
(1-t_2^2)}{t_1^2\,t_2^2}
\left[\varepsilon_1 \mu_1\,
\frac{t_1\,(1-t_2^2)}{t_2\,(1-t_1^2)}\, \left(-\frac{5}{128}\,
t_1\,t_2^2+\frac{41}{64}t_1\,t_2^4\right.\right.\right.\nn\\
&&-\frac{205}{384}\,t_1\,t_2^6+\frac{203}{1152}\,t_1^6\,t_2
-\frac{1015}{3456}\,t_1^6\,t_2^3-\frac{13}{64}\,t_1^4\,t_2+
\frac{65}{192}\,t_1^4\,t_2^3\nn\\
&&+\frac{1435}{3456}\,t_1^3\,t_2^6
+\frac{3}{128}\,t_1^2\,t_2-\frac{5}{128}\,t_1^2 \,t_2^3 
-\frac{21}{128}\,t_2^3 
+\frac{3671}{1920}\,t_2^5\nn\\
&&-
\frac{4543}{1152}\,t_2^7+\frac{22715}{10368}\,t_2^9
+\frac{35}{384}\,t_1^3\,t_2^2
-\frac{287}{576}\,t_1^3 \,t_2^4
+\frac{21385}{10368}\,t_1^9\nn\\
&&-\frac{15}{128}\,t_1^3-
\frac{469}{128}\,t_1^7 
\left.+\frac{1103}{640}\,t_1^5
\right)\nn\\
&&+(\varepsilon_1 \mu_2+\varepsilon_2 \mu_1)\, \left(
\frac{3}{128}\,t_1\,t_2^2-\frac{11}{192}\,t_1\,t_2^4+\frac{35}{1152}\,
t_1\,t_2^6-\frac{35}{1152}\,t_1^6\,t_2\right.\nn\\
&&+\frac{35}{3456}\,t_1^6 \,t_2^3
+\frac{11}{192}\,t_1^4\,t_2-\frac{11}{576}\,t_1^4\,t_2^3-
\frac{35}{3456}\,t_1^3 \,t_2^6-\frac{3}{128}\,t_1^2\,t_2\nn\\
&&+
\frac{1}{128}\,t_1^2\,t_2^3
+\frac{9}{128}\,t_2^3-\frac{293}{640}\,t_2^5
+\frac{789}{1152}\,t_2^7-\frac{3115}{10368}\,t_2^9-
\frac{1}{128}\,t_1^3\,t_2^2\nn\\
&&+\frac{11}{576}\,t_1^3\,t_2^4
\left.+\frac{3115}{10368}\,t_1^9-\frac{9}{128}\,t_1^3-\frac{787}{1152} 
\,t_1^7+\frac{293}{640}\,t_1^5\right)\nn\\
&&
+\varepsilon_2\,\mu_2 \frac{t_2\,(1-t_1^2)}{t_1\,(1-t_2^2)}
\left(-\frac{3}{128}\,t_1\,t_2^2+\frac{13}{64}\,t_1\,t_2^4-
\frac{203}{1152}\,t_1\,t_2^6+
\frac{205}{384} \,t_1^6\,t_2\right.\nn\\
&&-\frac{1435}{3456}\,t_1^6\,t_2^3-\frac{41}{64}\,t_1^4\,t_2+
\frac{287}{576}t_1^4\,t_2^3+
\frac{1015}{3456}\,t_1^3 \,t_2^6+\frac{15}{128}\,t_1^2\,t_2\nn\\
&& +\frac{35}{384}\,t_1^2\,t_2^3+\frac{15}{128}\,t_2^3
-\frac{1103}{640}\,t_2^5+\frac{469}{128}\,t_2^7 
-\frac{21385}{10368}\,t_2^9+\frac{5}{128}\,t_1^3\,t_2^2\nn\\
&&-\frac{65}{192}\,t_1^3\,t_2^4 
\left.\left.-\frac{22715}{10368}\,t_1^9+
\frac{21}{128}\,t_1^3+\frac{4543}{1152}\,t_1^7-
\frac{3671}{1920}\,t_1^5\right)\right] \nn \\
&&+(1-\eta^2)\,\eta^2\,(1-\alpha^2)^2
\varepsilon_2 \,\mu_2 \,t_1\,t_2\,
\left(\frac{5}{128}\,t_1\,t_2^2-\frac{41}{192}\,t_1\,t_2^4
+\frac{205}{1152}\,t_1\,t_2^6 \right. \nn \\
&&-\frac{205}{1152}\,t_1^6\,t_2+\frac{1025}{3456}
\,t_1^6\,t_2^3+
\frac{41}{192}\,t_1^4\,t_2-\frac{205}{576}\,t_1^4 \,t_2^3-
\frac{1025}{3456}\,t_1^3\,t_2^6\nn\\
&&-\frac{5}{128}\,t_1^2 \,t_2 +\frac{25}{384}\,t_1^2\,t_2^3-\frac{21}{128}\,t_2^3+
\frac{3671}{1920}\,t_2^5-\frac{4543}{1152}\,t_2^7+
\frac{22715}{10368}\,t_2^9\nn\\
&&-\frac{25}{384}\,t_1^3\,t_2^2+
\frac{205}{576}\,t_1^3\,t_2^4-\frac{22715}{10368}\,t_1^9
+\frac{21}{128}\,t_1^3 
+\frac{4543}{1152}\,t_1^7\nn\\
&&\left.\left.-\frac{3671}{1920}\,t_1^5\right)\right\},
\end{eqnarray}
and the denominator is
\begin{eqnarray}
{\cal F}&=&\frac{(1-t_1^2)\,
(1-t_2^2)}{t_1^2\,t_2^2}
\left[\varepsilon_1 \mu_1\,
\frac{t_1\,(1-t_2^2)}{t_2\,(1-t_1^2)}\, 
+
(\varepsilon_1 \mu_2+\varepsilon_2 \mu_1)\, 
+\varepsilon_2\,\mu_2 \frac{t_2\,(1-t_1^2)}{t_1\,(1-t_2^2)}
\right] \nn \\
&&+(1-\eta^2)\,\eta^2\,(1-\alpha^2)^2
\varepsilon_2 \,\mu_2 \,t_1\,t_2.
\end{eqnarray}


\bibliographystyle{unsrt}
\bibliography{/home/bordag/Literatur/Bordag,/home/bordag/Literatur/Gilkey,/home/bordag/Literatur/libri,/home/bordag/Literatur/articoli,/home/bordag/Literatur/Milton,/home/bordag/Literatur/Barton,/home/bordag/Literatur/Nesterenko,/home/bordag/Literatur/Marachevsky}

\end{document}